\documentclass[10pt, journal]{IEEEtran}
\ifCLASSINFOpdf

\else

\fi
\usepackage{amsfonts}
\usepackage{amsmath}
\usepackage{amsthm}
\usepackage{graphicx}
\usepackage{epstopdf}
\usepackage{etoolbox}
\usepackage{comment}
\usepackage[caption=false]{subfig}
\usepackage{cite}

\newtheorem{proposition}{Proposition}

\newcommand\floor[1]{\lfloor#1\rfloor}
\newcommand\ceil[1]{\lceil#1\rceil}

\begin{document}

\title{On Coded Caching in the Overloaded MISO Broadcast Channel}

\author{\IEEEauthorblockN{Enrico~Piovano,
		Hamdi~Joudeh
		and~Bruno~Clerckx} \\
	    \thanks{This work has been partially supported by the EPSRC of UK, under grant EP/N015312/1.}
	\IEEEauthorblockA{Department of Electrical and Electronic Engineering, Imperial College London,
		United Kingdom \\
		Email: \{e.piovano15, hamdi.joudeh10, b.clerckx\}@imperial.ac.uk}}

\maketitle

\begin{abstract}
	 This work investigates the interplay of coded caching and spatial
	 multiplexing in an overloaded Multiple-Input-Single-Output (MISO) Broadcast Channel (BC),
	 i.e. a system where the number of users is greater than the number
	 of transmitting antennas.
	 On one hand, coded caching uses the aggregate global cache memory of the users to create multicasting opportunities. On the other hand, multiple antennas at the transmitter leverage the available CSIT to transmit multiple streams simultaneously.
	 In this paper, we introduce a novel scheme which combines
	 both the gain derived from coded-caching and spatial multiplexing and outperforms existing schemes in terms of delivery time and CSIT requirement.
	
\end{abstract}


\IEEEpeerreviewmaketitle
\section{Introduction} \label{introduction}
Caching is a promising technique proposed to improve the throughput and reduce the latency in communication networks \cite{Golrezaei2012,maddahali_1,maddahali_2}.
In a seminal work by Maddah-Ali and Niesen  \cite{maddahali_1}, the fundamental limits of cache-aided networks were explored by considering a setting in which a single transmitter (server) communicates with multiple users over a shared medium. The analysis revealed that although users cannot cooperate, there exists a (hidden) global caching gain that scales with the aggregated memory distributed across the network, alongside the more obvious local caching gain.

In the context of wireless networks, recent efforts have been made to combine
coded caching with the conventional interference management techniques \cite{Gesbert2010,Jafar2011}
and proved to provide performance gains in different settings \cite{Maddah-Ali2015,Naderializadeh2016,elia_1,elia_2,kobayashi}.
In this work, we focus on the cache-aided multiple-input-single-output (MISO) broadcast channel (BC), in which a transmitter equipped with multiple antennas serves multiple single-antenna users equipped with cache memories.

\emph{Overloading the MISO BC:}
The ability to simultaneously serve a large number of users is a key feature envisioned for future wireless networks,
pushing them towards overloaded regimes in which the number of users exceeds the number of transmit antennas.
In this context, the setting considered in \cite{maddahali_1} is intrinsically an overloaded BC where all nodes are equipped with a single antenna.
In this work, we consider a more general overloaded BC in which the transmitter is allowed to have multiple antennas $K>1$, which is yet still smaller than the number of users $K_{\mathrm{t}}$.
A recent work on overloaded system can be found in \cite{Piovano2016}.
In this paper, we make progress towards characterizing the optimum caching strategy and cache-aided performance in such setting, which remain unknown.
To simplify the analysis, we consider scenarios in which the number of scheduled users is an integer multiple of the number of antennas, i.e. $K_{\mathrm{t}} = GK$, where $G$ is a positive integer denoted as an overloading factor.
First, we consider the above setting under the assumption of perfect CSIT.
A natural way to serve the $GK$ users is to divide them into $G$ groups of $K$ users, each served independently in an orthogonal manner (e.g. over time). 
Caching is carried out independently for each group in which only the local caching gain is relevant.
This is denoted by the Orthogonal Scheme (OS).
An alternative scheme is the one proposed by Maddah-Ali and Niesen (MAN) \cite{maddahali_1},
in which spatial multiplexing gains from zero-forcing are completely ignored.
We show that both strategies are in fact suboptimal for the considered setting by proposing a scheme that outperforms both in terms of the \textit{delivery time}, i.e. the time required to deliver the requested information during the \textit{delivery phase}.
\emph{Proposed scheme (PS):}
We propose to partition the library of files at the transmitter into two parts by dividing each file into two subfiles. One part of the library is stored in the global memory of all $K_{\mathrm{t}}$ users in the MAN manner (cached part), and the remaining part is never cached (uncached part).
During the \emph{delivery phase}, information requested from the cached part is transmitted through a single coded multicasting stream.
This is superposed on top of a zero-forcing layer shared in an orthogonal manner between the $G$ groups, and carrying information requested from the uncached part.
From its structure, it can be seen that such scheme can exploit both the zero-forcing gains of the MISO BC, and the global cache memory of all users.
We show that under adequate partitioning of the library, PS outperforms both OS and MAN.
\emph{Partial CSIT:}
We relax the assumption of perfect CSIT.
We show that the \textit{delivery time} achieved by the PS with perfect CSIT is in fact maintained under partial CSIT, up to a certain quality.
We then extend the PS to deal with any partial CSIT level. This is implemented by caching a fraction of the library tailored to the actual CSIT and we show, through numerical results, the gain compared to the MAN scheme. We then compare with the OS.  However, we consider now the coded caching strategy in \cite{elia_1} applied independently to each group of $K$ users. We show that this allows to serve the overloaded system by achieving the same \textit{delivery time} as in perfect CSIT with a reduced quality.
We prove then that the PS achieves the same \textit{delivery time} with a further reduced CSIT requirement.

%
\section{System Model} \label{system_model}

\subsection{Overloaded cache-aided MISO BC} \label{SM_A}

We consider the overloaded scenario described before. The transmitter has access to a library with $N_{\mathrm{f}}$ distinct files, denoted as $W_1, W_2, \dots, W_{N_{\mathrm{f}}}$, each of size $f$ bits and we assume $N_{\mathrm{f}} \geq K_{\mathrm{t}}$. Each user $k \in \mathcal{K}_{\mathrm{t}}$ is equipped with a cache-memory $Z_k$ of size $Mf$ bits, where $M < N_{\mathrm{f}}$.
For tractability, we assume that the ratio $\Gamma=\frac{K_{\mathrm{t}}M}{N_{\mathrm{f}}}$ is an integer. This allows for a closed-form expression of the achievable delivery time for the MAN scheme.
The communication takes place over two phases: the \textit{placement phase} and the \textit{delivery  phase}. During the \textit{placement phase}, before actual user demands are revealed, the caches $Z_k$ are pre-filled with information from the $N_{\mathrm{f}}$ files $W_1, W_2, \dots, W_{N_{\mathrm{f}}}$.
During the \textit{delivery phase}, each user $k$ requests a single file $F_k$, for some $F_k \in \{1,2,\cdots,N_{\mathrm{f}}\}$.
The requested files $W_{F_{1}},\ldots,W_{F_{K_{\mathrm{t}}}}$ are jointly mapped  into the transmitted signal
 $\mathbf{x} \in \mathbb{C}^{K \times 1}$,  which satisfies the power constraint $\mathbb{E}(\mathbf{x}^H \mathbf{x}) \leq P $.
At the $t$-th discrete channel use, the $k$-th user's received signal is given by
\begin{equation}
{y}_k(t)=\mathbf{h}_k^H(t)\mathbf{x}(t)+n_k(t),\quad k=1,2,\dots,K_{\mathrm{t}}
\end{equation}
 where $\mathbf{h}^H_k(t) \in \mathbb{C}^{1 \times K}$ is the channel between the transmitter and the $k$-th user and $n_k(t) \sim \mathcal{CN}(0,1)$ is the Additive White Gaussian Noise (AWGN).
 The duration of the \textit{delivery phase} is given by $T$, which is a normalized measure as explained in the next
 subsection. The discrete channel uses are indexed by $t$, where $t \in [0,T]$ as in \cite{elia_2}. In the remainder of the paper, the index $t$ of the channel use will be omitted for ease of notation.
 At the end of the \textit{delivery phase}, each user combines the signal $y_k$ with its own cached information $Z_k$ in order to retrieve the requested file $W_{F_k}$.

\subsection{Performance measure}

In this paper, the analysis is restricted to the high-SNR regime. Such analysis gives insight into the role of coded caching in interference management.
The metric of evaluation is the duration $T$ needed  to complete the \textit{delivery phase} for every request by the users. $T$ is measured in time slots per file served. As in \cite{elia_1}, the time is normalized such that one time slot is the amount of time required to transfer a single file to a single receiver without caching and interference.
 As a consequence, since the single-stream capacity scales as $\log_2(P)$ for the high-SNR regime, we can impose the size of each file $f=\log_2(P)$ to align our measure of performance with \cite{maddahali_1} and make the two comparable.
Finally, it is worth noting that similar to \cite{maddahali_1,elia_1}, $T$ represents the worst case scenario, i.e. the case when each user
requests a different file.
For ease of notation, and without loss of generality, we assume that user $k$
requests file $W_k$ for all $k \in \mathcal{K}_{\mathrm{t}}$.
\subsection{Partial Instantaneous CSIT}
To study the influence of CSIT imperfections, the channel is modelled by
 \begin{equation}
 {\mathbf{h}}_k={\hat{\mathbf{h}}}_k+\tilde{\mathbf{h}}_k
 \end{equation}
 where $\hat{\mathbf{h}}_k$ denotes the instantaneous channel estimate available at the transmitter for user $k$, and ${\tilde{\mathbf{h}}}_k$ denotes the CSIT error, which is assumed to have a covariance matrix $\sigma_k^2 \mathbf{I}$.
 The variance $\sigma_k^2$ is parametrized as a function of the SNR $P$ in the form of $P^{-\alpha_k}$, where $\alpha_k$ is the \textit{CSIT quality exponent} defined as
 \begin{equation}
 \alpha_k = \lim_{P \to \infty} {-\frac{\log(\sigma_k^2)}{\log(P)}},\quad k=1,2,\dots,K_t.
 \end{equation}
The exponent is restricted to $\alpha_k \in [0,1]$ which captures the entire range of CSIT (from unknown to perfect) in the high-SNR regime  \cite{jindal,Yang2013}.

%
 \section{Multiplexing and Multicasting} \label{problem overview}

In this section, we present what we consider to be the two most obvious ways to deal with the cached-aided scenario at hand. We further assume that CSIT is perfectly known.
\subsection{The Orthogonal Scheme (OS)}
 The OS transmits $K$ interference-free streams due to the presence
 of $K$ antennas and perfect CSIT.
 Each user stores the same fraction from each file in the library, hence during the \textit{placement phase} its memory is filled with $\frac{M}{N_{\mathrm{f}}}f$ bits from each file.  This leaves $(1-\frac{M}{N_{\mathrm{f}}})f$ bits to be delivered during the \textit{delivery phase}.
 The  $K_{\mathrm{t}}$ users are divided into $G$ groups of $K$ users each, where groups are orthogonalized in time
 and each group is served using zero-forcing. It follows that the \textit{delivery time} of each group is given
 by $1-\frac{M}{N_{\mathrm{f}}}$ from which the total \textit{delivery time} is given by
   \begin{equation} \label{OR_TD}
   T_{\mathrm{D}}^{\mathrm{OS}}=G\left(1-\frac{M}{N_{\mathrm{f}}}\right).
   \end{equation}
 The term $1-\frac{M}{N_{\mathrm{f}}}$ corresponds to the local caching gain achieved due to the locally stored content
 in the memory of each user.
It is evident that the OS only exploits the local caching gain and no coded multicasting is needed.
In fact multiuser interference is eliminated through time-sharing and zero-forcing.
 \subsection{Maddah-Ali Niesen scheme (MAN)}
Alternatively, all $K_{\mathrm{t}}$ users can be served jointly using the MAN scheme proposed in \cite{maddahali_1}. While this scheme is designed for a system with a single transmitting antenna, it can be used here by ignoring the multiplexing gains of the antenna array. Following the same placement and delivery procedure in \cite{maddahali_1}, the total \textit{delivery time} is given by
\begin{equation} \label{MA_TD}
T_{\mathrm{D}}^{\mathrm{MAN}}=K_{\mathrm{t}}\left(1-\frac{M}{N_{\mathrm{f}}}\right)\frac{1}{1+\Gamma} \cdot
\end{equation}
The term $1-\frac{M}{N_{\mathrm{f}}}$ corresponds to the local caching gain as in the OS, while the global caching gain is captured by $\frac{1}{1+\Gamma} \cdot$

\subsection{Motivations for a new scheme}
We compare the performance of OS and MAN using the ratio $\frac{T^{\mathrm{OS}}_{\mathrm{D}}}{T^{\mathrm{MAN}}_{\mathrm{D}}}=\frac{1+\Gamma}{K}$.
The value of $\Gamma$, which is an integer, belongs to the set $\{1,\ldots,K_{\mathrm{t}}\}$.
It is evident from the ratio
that the OS performs comparatively better or equal to the MAN scheme for $\Gamma \in \{1,\ldots,K-1\}$. On the other hand, for $\Gamma \geq K$, MAN scheme performs better than the OS scheme.
The OS and MAN schemes can be seen as the two extreme cases of utilizing the available CSIT on one side and caching on the other side.
We propose a scheme that combines elements of OS and MAN by incorporating both spatial multiplexing and global cache gains.
This is shown to outperform both  OS and MAN in the region $\Gamma \in \{1,\ldots,\zeta\}$, where $\zeta \geq K-1$ is some integer.
For the region $\Gamma \in \{\zeta+1, \ldots, K_{\mathrm{t}}\}$, the proposed scheme matches the performance of MAN.
In this paper we focus on $\Gamma \leq K-1$, i.e. $\Gamma$ belongs to the set $\{1,\ldots,K-1\}$.

\section{Proposed Scheme} \label{proposed scheme}
The design of the PS involves a partition of the library into two parts: one treated in the MAN manner, while the other is never cached, hence delivered using zero-forcing.
Such partition is implemented by caching a fraction $p \in [0,1]$ of each file, so that the MAN scheme consists of $p N_{\mathrm{f}}$ files, while the remaining $(1-p) N_{\mathrm{f}}$ files correspond to the zero-forcing part.
Since we treat the cached part in a MAN manner, the ratio  $\eta=\frac{K_{\mathrm{t}}M}{pN_{\mathrm{f}}}$ is assumed to be an integer, where $\eta$ indicates the number of times the cached part is repeated inside the memories.
In case of $p=1$, we cache all the library and $\eta=\Gamma$.
On the other hand, for $p=\frac{M}{N_{\mathrm{f}}}$, the cached part fits exactly in every single memory and in this case $\eta=K_{\mathrm{t}}$.

We can write $p$ in terms of $\eta$, i.e.
$p=\frac{K_{\mathrm{t}}M}{\eta N_{\mathrm{f}}}$, where $\eta \in [\Gamma,K_{\mathrm{t}}] \cap \mathbb{Z}$.
Increasing $\eta$ translates into a decreased cached portion from each file and higher replication in the memories which, in turn, is translated into multicasting messages intended for more receivers. However, a larger uncached part needs to be delivered using  zero-forcing.
Selecting the right parameter $\eta$ (or equivalently the right $p$), is crucial to achieve an optimum trade-off between spatial multiplexing and coded multicasting gains as we see in what follows.
Next, the placement and delivery phase of the PS is described in detail.

 \subsection{Placement Phase} \label{PP_1}
We first describe the placement and delivery for a given factor $\eta$, or equivalently $p=\frac{K_{\mathrm{t}}M}{\eta N_{\mathrm{f}}}$. Then $\eta$ is optimized to minimize the \textit{delivery time}.
We divide each of the $N_{\mathrm{f}}$ files $W_1, W_2, \dots, W_{N_{\mathrm{f}}}$ into two subfiles:
\begin{equation}
W_n=(W_n^{(\mathrm{c})},W_n^{(\mathrm{p})})
\end{equation}
where $W_n^{(\mathrm{c})}$, of size $pf$, is cached into the memories while $W_n^{(\mathrm{p})}$, of size $(1-p)f$, is never cached.
We apply the idea in \cite{maddahali_1} to the subfiles $W^{\mathrm{(c)}}_1, W^{\mathrm{(c)}}_2, \dots, W^{\mathrm{(c)}}_{N_{\mathrm{f}}}$ by jointly encoding all $K_{\mathrm{t}}$ users together considering their total {global cache memory}.
Consequently, according to \cite{maddahali_1}, each subfile $W_n^{(\mathrm{c})}$ is split into $K_{\mathrm{t}} \choose \eta$ smaller subfiles $W^{\mathrm{(c)}}_{n, \tau}$, for all $\tau \in \Omega$, where $\Omega=\{ \tau \subset \mathcal{K}_{\mathrm{t}} : |\tau|=\eta \}$. It follows that each subfile $W^{\mathrm{(c)}}_{n, \tau}$ has size $\frac{pf}{{K_{\mathrm{t}} \choose \eta}}$.
From each file $n \in \{1,2,\dots,N_{\mathrm{f}}\}$, user $k$ caches the subfiles $W^{\mathrm{(c)}}_{n, \tau}$
such that $k \in \tau$.
Hence each user caches $\frac{Mf}{N_{\mathrm{f}}}$ bits from any $W_n^{(\mathrm{c})}$.
Since we assume different demands across users, the total number of files to be delivered from the
cached part of the library is given by
$Q_{\eta}^{(\mathrm{c})}=K_{\mathrm{t}}({\frac{K_{\mathrm{t}}M}{\eta N_{\mathrm{f}}}-\frac{M}{N_{\mathrm{f}}}})\frac{1}{1+\eta} \cdot$
Additionally, the number of files to be delivered through zero-forcing is given by $Q_{\eta}^{\mathrm{(p)}}=K_{\mathrm{t}}\left(1-\frac{K_{\mathrm{t}}M}{\eta N_{\mathrm{f}}}\right) \cdot$
\subsection{Delivery Phase} \label{delivery phase}

During the \textit{delivery phase}, the actual demands from users are revealed.
A given file $W_{k}$ requested by the corresponding user $k$ consists of three parts: a) a part that is contained in the cache $Z_k$ and hence it is not requested during the \textit{delivery phase}, b) a part that is contained in the memories of other users $Z_{m}$ for $m \in \mathcal{K}_{\mathrm{t}} \setminus k$, which is termed as the requested cached part and given by the subfiles $W^{\mathrm{(c)}}_{k, \tau}$, for all $k \notin \tau$ c) a part that has never been cached termed as the uncached part which is denoted by ${W}_k^{(\mathrm{p})}$.
The \textit{delivery phase} is carried out using a superposed transmission of the requested cached part, where the coded multicasting information is delivered by a common symbol decoded by all $K_t$ users, and
the uncached parts, delivered through private symbols in a zero-forcing fashion.
Since the users share the zero-forcing layer in a orthogonal manner, the transmission is carried out over $G$ sub-phases indexed by $g=1,\dots,G$.
Without loss of generality, we assume that in the $g$-th sub-phase, the private symbols are intended to the group $\mathcal{K}_{\mathrm{g}}=\{(g-1)K+1,\ldots,gK\}$.

We denote the common symbols transmitted over the $G$ sub-phases as $\{x_g^{\mathrm{(c)}}\}_{g=1}^{G}$.
They carry the coded messages
\begin{equation} \label{common_message}
W_{\mathcal{S}}^{(\mathrm{c})}=(\oplus_{k \in \mathcal{S}}W^{(\mathrm{c})}_{k, \mathcal{S} \setminus \{k\}} : \mathcal{S} \in \Theta)
\end{equation}
where $\Theta=\{\mathcal{S} \subset \mathcal{K}_{\mathrm{t}} : |\mathcal{S}|= \eta+1\}$.  These coded messages are a combination
of the requested cached part as they were first proposed in \cite{maddahali_1} and each of them contains information for $\eta+1$ users.
Similarly, at the $g$-th sub-phase, private symbols $x^{\mathrm{(p)}}_{k}$ carry the uncached parts $W^{\mathrm{(p)}}_{k}$, where $k \in \mathcal{K}_g$.
During the $g$-th delivery sub-phase, the transmitted signal is
\begin{equation}
\mathbf{x}_g= \underbrace{\sqrt{P}\mathbf{v}^{(\mathrm{c})}x_{g}^{(\mathrm{c})}}_{\mathrm{MAN}} +\underbrace{\sum_{k \in \mathcal{K}_g} \sqrt{P^{\beta}} \mathbf{v}^{(\mathrm{p})}_{k} x_k^{(\mathrm{p})}}_{\mathrm{ZF}} +
O(1)
\end{equation}
where $\mathbf{v}^{(\mathrm{c})} \in \mathbb{C}^{M \times 1}$ and $\mathbf{v}^{(\mathrm{p})}_{k} \in \mathbb{C}^{M \times 1}$ are unitary precoding vectors, and $\beta \in [0,1]$ is the power partitioning factor \cite{Piovano2016}.
Note that the term $O(1)$ guarantees that the power constraint is not violated. Such term has no influence on high-SNR analysis.
Since $x^{\mathrm{(c)}}_{g}$ has to be decoded by all $K_{\mathrm{t}}$ users, $\mathbf{v}^{(\mathrm{c})}$ is chosen as a generic precoding vector.
On the other hand, private symbols are precoded using the well-known zero-forcing precoders.

 The received signal for each user $k \in \mathcal{K}_{\mathrm{t}}$ is given by
\begin{equation*} \label{prop_rx_signal}
y_k= \sqrt{P} {\mathbf{h}_{k}^H}
\mathbf{v}^{(\mathrm{c})}x_{g}^{(\mathrm{c})}
 +\sum_{i \in \mathcal{K}_g} \sqrt{P^{\beta}} {\mathbf{h}_{k}^H} \mathbf{v}^{(\mathrm{p})}_{i} x_i^{(\mathrm{p})} +
 O(1).
\end{equation*}
All users in $\mathcal{K}_{\mathrm{t}}$ decode $x^{(\mathrm{c})}_{g}$ by treating the interference from all the other symbols as noise. Hence, $x^{\mathrm{(c)}}_{g}$ can carry
$(1-\beta)\log_2(P)$ bits per time slot.
Users in $\mathcal{K}_g$ proceed to remove the contribution of $x^{\mathrm{(c)}}_{g}$ to the received signal and decode their private symbols $x^{(\mathrm{p})}_{k}$.
The symbol $x^{(\mathrm{p})}_{k}$ is decoded  with a SINR of $O(P^{\beta})$, hence carrying $\beta\log_2(P)$ bits per time slot.
It follows that the cached part of the library is delivered with a rate of  $(1-\beta)\log_2(P)$ bits per time slot and the uncached part of the library with a rate of $K\beta\log_2(P)$ bits per time slot.
From $\{x^{\mathrm{(c)}}_{g}\}_{g=1}^{G}$ each user $k \in \mathcal{K}_{\mathrm{t}}$ retrieves the messages $W_{\mathcal{S}}^{(\mathrm{c})}$ and consequently its cached part.
In fact any missing sub-files $W^{\mathrm{(c)}}_{k, \tau}$, so that $k \notin \tau$, can be  recovered by combining $W_{\mathcal{S}}^{(\mathrm{c})}$ with the pre-stored sub-files
$W^{\mathrm{(c)}}_{i, \mathcal{S} \setminus \{i\}}$, where $\mathcal{S}= \tau \cup \{k\}$ and $i \in \mathcal{S} \setminus \{k\}$.
On the other hand, from  $x^{(\mathrm{p})}_{k}$ each user $k \in \mathcal{K}_{\mathrm{g}}$ retrieves the uncached part ${W}_k^{(\mathrm{p})}$.
\subsection{Delivery time} \label{PCSIT_delivery_time}
During the communication both the cached and uncached parts must be delivered.
We indicate as $T^*_{\mathrm{D},\eta}$ the \textit{delivery time}
of such communication.
The power partitioning factor, indicated as $\beta^*$, is chosen to let both parts to be delivered over this \textit{delivery time}.
Hence, the two equalities $Q_{\eta}^{\mathrm{(p)}}f = K\beta^* f T^*_{\mathrm{D},\eta}$ and
$Q_{\eta}^{\mathrm{(c)}}f = (1-\beta^*)fT^*_{\mathrm{D},\eta}$ must be simultaneously satisfied.
It can be verified that the solution is given by
$T^*_{\mathrm{D},\eta}=\frac{Q_{\eta}^{\mathrm{(p)}}}{K}+Q_{\eta}^{\mathrm{(c)}}$
and $\beta^*=\frac{Q_{\eta}^{\mathrm{(p)}}/K}{Q_{\eta}^{\mathrm{(p)}}/K+Q_{\eta}^{\mathrm{(c)}}}$.
By expanding $Q_{\eta}^{\mathrm{(c)}}$ and $Q_{\eta}^{\mathrm{(p)}}$,  we obtain
\begin{equation} \label{T_d,eta}
    T^*_{\mathrm{D},\eta}=G\left(1-\frac{MK_{\mathrm{t}}}{\eta  N_{\mathrm{f}}}\right)+\frac{K_{\mathrm{t}}}{\eta+1} \left( \frac{MK_{\mathrm{t}}}{\eta N_{\mathrm{f}}} -\frac{M}{N_{\mathrm{f}}} \right).
\end{equation}
The value of $\eta \in [\Gamma,K_{\mathrm{t}}] \cap \mathbb{Z}$ must be chosen to minimize the \textit{delivery time} in (\ref{T_d,eta}).
This is given by
\begin{equation} \label{eq_T}
T^*_{\mathrm{D}}= \min_{\eta \in [\Gamma,K_{\mathrm{t}}] \cap \mathbb{Z}} T^*_{\mathrm{D}, \eta}.
\end{equation}
where $T^*_{\mathrm{D}}$ denotes the minimum \textit{delivery time}.
The following result holds for which proof can be found in the
appendix.
\begin{proposition} \label{theo_1}
	In case of perfect CSIT, the PS achieves a delivery time $T^*_{\mathrm{D}}$ given by
   \begin{equation} \label{UP}
   T^*_{\mathrm{D}}= \min_{\eta \in \{\floor{x},\ceil{x} \} } T^*_{\mathrm{D}, \eta}
   \end{equation}
 where
 \begin{equation} \label{x}
   x=\frac{G(K-1)+\sqrt{G^2(K-1)^2+G(G+1)(K-1)}}{G+1}.
\end{equation}
\end{proposition}
The value of $\eta$ that minimizes the \textit{delivery time} in (\ref{UP}) is denoted by $\eta^{*}$, i.e. $T^*_{\mathrm{D}}= T^*_{\mathrm{D},\eta^*}$.
By comparing the result in Preposition \ref{theo_1} with the \textit{delivery time} in (\ref{OR_TD}) and (\ref{MA_TD}), it is verified that $  T^*_{\mathrm{D}} <   T^{OS}_{\mathrm{D}} \leq   T^{\mathrm{MAN}}_{\mathrm{D}}$.
 Hence, simultaneously exploiting the global cache memory of all users and the multiplexing gain, the PS strictly outperforms both OS and MAN.
\section{Partial CSIT} \label{imperfect CSIT}
In this section, we relax the constraint of perfect CSIT and consider a setting with partial CSIT. The scheme in turn
is generalized to account for such imperfection.
We start by showing that the \textit{delivery time} in (\ref{UP}) can be maintained, while relaxing the CSIT quality
$\alpha$. Then we modify the PS to be adaptable for any $\alpha$, by properly
tailoring the cached part of the library based on the available CSIT.
\subsection{CSIT Relaxation} \label{csit_relaxation}
In Section \ref{delivery phase}, perfect CSIT was exploited to deliver $K$ interference-free private symbols,
transmitted with a power which scales as $O(P^{\beta^*})$, where $\beta^*=\frac{Q_{\eta^*}^{\mathrm{(p)}}/K}{Q_{\eta^*}^{\mathrm{(p)}}/K+Q_{\eta^*}^{\mathrm{(c)}}}$.
If the channel estimation error decays as $O(P^{-\alpha})$, where \mbox{$\alpha \geq \beta^*$}, each user can still decode its own private symbol as interference from other symbols is at (or below) the noise floor. \cite{Joudeh2016}. Hence the SINR still scales as in case of perfect CSIT.
It follows that, the \textit{delivery time} $T_{\mathrm{D}}^{*}$ is mantained for any CSIT quality no less than $\beta^*$. Since this represents a CSIT quality threshold, by indicating it as $\alpha^*$, we obtain
\begin{equation} \label{alpha_th}
\alpha^*=\beta^*=\frac{G\left(1-\frac{MK_{\mathrm{t}}}{\eta^*  N_{\mathrm{f}}}\right)}{G\left(1-\frac{MK_{\mathrm{t}}}{\eta^*  N_{\mathrm{f}}}\right)+\frac{K_{\mathrm{t}}}{\eta^*+1} \left( \frac{MK_{\mathrm{t}}}{\eta^* N_{\mathrm{f}}} -\frac{M}{N_{\mathrm{f}}} \right)}.
\end{equation}
\subsection{General Case}
 The case \mbox{$\alpha < \alpha^*$} is treated by caching a larger fraction of the library which reduces the  load carried by the private symbols.
We proceed by considering a given $\eta$ which will be then optimized based on the actual CSIT.
For the delivery of the uncached part, zero-forcing is replaced by rate-splitting, which is well suited to operate under partial CSIT \cite{Joudeh2016}.
The \textit{placement phase} and \textit{delivery phase}  follow the same steps as in perfect CSIT.
In rate-splitting, at the $g$-th delivery sub-phase, the uncached parts $W^{\mathrm{(p)}}_{k}$ with $k \in \mathcal{K}_g$ are divided and partially mapped to the private symbols $x^{\mathrm{(p)}}_{k}$, while the remaining parts are jointly encoded into the symbol $\tilde{x}_g$, decoded by all users in $\mathcal{K}_g$.
 The transmitted signal is
\begin{equation*} \label{or_tx_signal}
\mathbf{x}_g= \underbrace{\sqrt{P}\mathbf{v}^{(\mathrm{c})}x_{g}^{(\mathrm{c})}}_{\mathrm{MAN}}+\underbrace{\sqrt{P^{\beta}}\mathbf{v}^{(\mathrm{c})}\tilde{x}_{g} +\sum_{k \in \mathcal{K}_g} \sqrt{P^{a}} \mathbf{v}^{(\mathrm{p})}_{k} x_k^{(\mathrm{p})}}_{\mathrm{RS}} + O(1)
\end{equation*}
where symbols, precoding vectors and powers are defined as in the previous section, with the difference that the power of the private symbols scales as $O(P^{a})$, with $a \leq \beta$.
Furthermore, to force interference between private symbols to the noise floor level, we need $a \leq \alpha$.
Hence, we take $a=\min(\alpha,\beta)$.
All users in $\mathcal{K}_{\mathrm{t}}$ decode $x_{g}^{(\mathrm{c})}$ by treating the other symbols as noise, hence it can carry $(1-\beta)\log_2(P)$ bits per time slot.
Then, all users in $\mathcal{K}_g$ proceed to cancel $x_{g}^{(\mathrm{c})}$ and decode $\tilde{x}_g$, which is then canceled before decoding their respective private symbols. It can be easily verified that rate-splitting achieves a total rate of $(\beta+(K-1)\min(\alpha,\beta))\log_2(P)$ bits per time slot.
 Further details on the rate-splitting procedure can be found in \cite{Joudeh2016,Piovano2016}.
As before, from $\{x^{\mathrm{(c)}}_{g}\}_{g=1}^{G}$ each user can retrieve its desired cached part, while
from $\tilde{x}_{g}$ and $x^{\mathrm{(p)}}_{k}$ users $k \in \mathcal{K}_g$ retrieve their uncached parts.
\subsection{Delivery Time} \label{imperfect_delivery_time}
As in Section \ref{PCSIT_delivery_time}, the power partitioning factor $\beta^*$ is chosen such that
\mbox{ $Q_{\eta}^{\mathrm{(p)}}f =(\beta^*+(K-1)\min (\alpha,\beta^*)) f T_{\mathrm{D},\eta}(\alpha)$} and
$Q_{\eta}^{\mathrm{(c)}}f = (1-\beta^*)fT_{\mathrm{D},\eta} (\alpha)$, where $T_{\mathrm{D},\eta}(\alpha)$ denotes the \textit{delivery time}.
The \textit{delivery time} is given by
\begin{equation} \label{formula_1}
	T_{\mathrm{D},\eta}(\alpha)=\begin{cases}
	T_{\mathrm{D},\eta}^*, \; \alpha \geq \alpha_{\eta}^{*}\\
	\frac{Q_{\eta}^{(\mathrm{p})}+Q_{\eta}^{(\mathrm{c})}}{1+(K-1)\alpha} , \; \alpha < \alpha_{\eta}^{*}
	\end{cases}
\end{equation}
where $\alpha_{\eta}^{*}=\frac{Q_{\eta}^{\mathrm{(p)}}/K}{Q_{\eta}^{\mathrm{(p)}}/K+Q_{\eta}^{\mathrm{(c)}}}$ and $T_{\mathrm{D},\eta}^{*}=\frac{Q_{\eta}^{\mathrm{(p)}}}{K}+Q_{\eta}^{\mathrm{(c)}}$.
It can be seen that given a certain partition $\eta$, the same \textit{delivery time} as in perfect CSIT can be maintained  as long as $\alpha \geq \alpha_{\eta}^*$.
On the other hand, in case of $\alpha < \alpha_{\eta}^*$, the \textit{delivery time} is given by the ratio between the total number of transmitted files and the total communication rate.
In case of $\eta=\eta^*$, we reduce to  $\alpha^*_{\eta^*}=\alpha^*$ and $T^*_{\mathrm{D},\eta^*}=T_{\mathrm{D}}^*$, in agreement with Section \ref{csit_relaxation}.
For a given quality $\alpha$, the minimum achievable \textit{delivery time}, denoted by $T_{\mathrm{D}}(\alpha)$,  is given as
\begin{equation} \label{eq_T_alpha}
T_{\mathrm{D}}(\alpha)= \min_{\eta \in [\Gamma,K_{\mathrm{t}}] \cap \mathbb{Z}} T_{\mathrm{D}, \eta}(\alpha).
\end{equation}
The following result holds for which the proof is given in the appendix.
\begin{proposition} \label{theo_2}
	Given a CSIT quality $\alpha$, we can achieve the following delivery time:
	\begin{equation} \label{f_theo_2}
	T_{\mathrm{D}}(\alpha)=\begin{cases}
	T_{\mathrm{D}}^{*}, \; \alpha \geq \alpha^{*} \\
	\min \{T_{\mathrm{D},\eta}^{*}, T_{\mathrm{D},\eta+1}^{*} \frac{1+(K-1)\alpha_{\eta+1}^*}{1+(K-1) \alpha} \}, \; \alpha < \alpha^{*}
	\end{cases}
	\end{equation}		
	where
	\begin{equation} \label{eta_theo_2}
	\eta= \arg \max_{\eta' \in [\Gamma,\eta^*) \cap \mathbb{Z}} \{\eta':\alpha_{\eta'}^* \leq \alpha \}.
	\end{equation}	
\end{proposition}
For $\alpha \geq \alpha^*$ we can achieve the same \textit{delivery time} as perfect CSIT given in Proposition \ref{theo_1}.
Examining the \textit{delivery time} from the expression in (\ref{f_theo_2}) for the entire range of $\alpha$ is not an easy task as we were not able to derive a more tractable expression.
However, it can be seen that for $\alpha = 0$, we obtain $\eta=\Gamma$ in (\ref{eta_theo_2}) and the scheme reduces to MAN.
The influence of increasing $\alpha$ is examined through simulations in the following section where it is shown that the PS leverages the available CSIT and outperforms the MAN scheme.

Looking at the PS from a different perspective, we investigate the minimum CSIT quality required to achieve a given delivery time $T$. This is given through the following result for which the proof is given in the appendix.
\begin{proposition} \label{proposition_2}
	A delivery time $T \geq T_{\mathrm{D}}^*$ can be achieved by a CSIT quality $\alpha$ equal to
	\begin{equation} \label{f_proposition_2}
	\alpha= \frac{T^{*}_{\mathrm{D}, \eta}\left[1+(K-1)\alpha^*_{\eta}\right]-T}{(K-1)T}
	\end{equation}		
	where
	\begin{equation} \label{eta_T_alpha}
	\eta= \arg \min_{\eta' \in [\Gamma ,\eta^*]  \cap \mathbb{Z}} \{\eta':T_{\mathrm{D},\eta'}^* \leq T \}.
	\end{equation}
\end{proposition}
Since the MAN scheme does not require CSIT, it is suitable to compare to the OS in this context. We further assume that each orthogonal group of $K$ users is served using the strategy in \cite{elia_1} to account for imperfect CSIT.
It follows that the \textit{delivery time} $T_{\mathrm{D}}^{\mathrm{OS}}$ can be achieved while applying coded caching and relaxing the CSIT quality to  \mbox{$\label{alpha_th_or_1}	\alpha_{\mathrm{th}}^{\mathrm{OS}}=\frac{N_{\mathrm{f}}-\frac{N_{\mathrm{f}}}{K}-M}{\frac{M}{K}+N_{\mathrm{f}}-\frac{N_{\mathrm{f}}}{K}-M}$}.
In comparision, by applying Proposition \ref{proposition_2}, we can characterize the minimum CSIT quality needed by the PS to achieve $T^{\mathrm{OS}}_{\mathrm{D}}$, denoted as $\alpha_{\mathrm{th}}^{\mathrm{PS}}$.
We observe that $T_{\mathrm{D}}^{\mathrm{OS}}=T_{\mathrm{D},K-1}^*$. Hence the value of  $ \alpha_{\mathrm{th}}^{\mathrm{PS}}$ in (\ref{f_proposition_2}) is given by
\begin{equation} \label{alpha_th_ps_1}
 \alpha_{\mathrm{th}}^{\mathrm{PS}}=\frac{N_{\mathrm{f}}-\frac{N_{\mathrm{f}}}{K}-MG}{\frac{M}{K}+N_{\mathrm{f}}-\frac{N_{\mathrm{f}}}{K}-M}.
\end{equation}
By comparing $ \alpha_{\mathrm{th}}^{\mathrm{OS}}$ and $ \alpha_{\mathrm{th}}^{\mathrm{PS}}$, we can see that $ \alpha_{\mathrm{th}}^{\mathrm{PS}}$ is strictly lower for the overloaded case ($G \geq 2$).
\section{Numerical results} \label{simulation_results}
\begin{figure}
 	\centering
 	\includegraphics[width=0.25\textwidth]{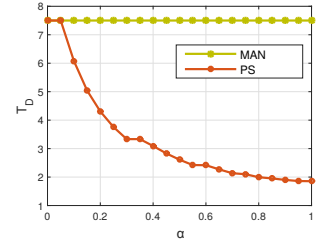}
 	\caption{$T_{\mathrm{D}}$ as function of the CSIT quality $\alpha$ for $K=8$ and $G=2$}
 	\label{fig:figura_0}
 \end{figure}

\begin{figure}
	\subfloat[$\alpha_{\mathrm{th}}^{\mathrm{PS}}$ and $\alpha_{\mathrm{th}}^{\mathrm{OS}}$ with respect to $G$ \label{sfig:figura_1a}]{%
		\includegraphics[width=.50\linewidth]{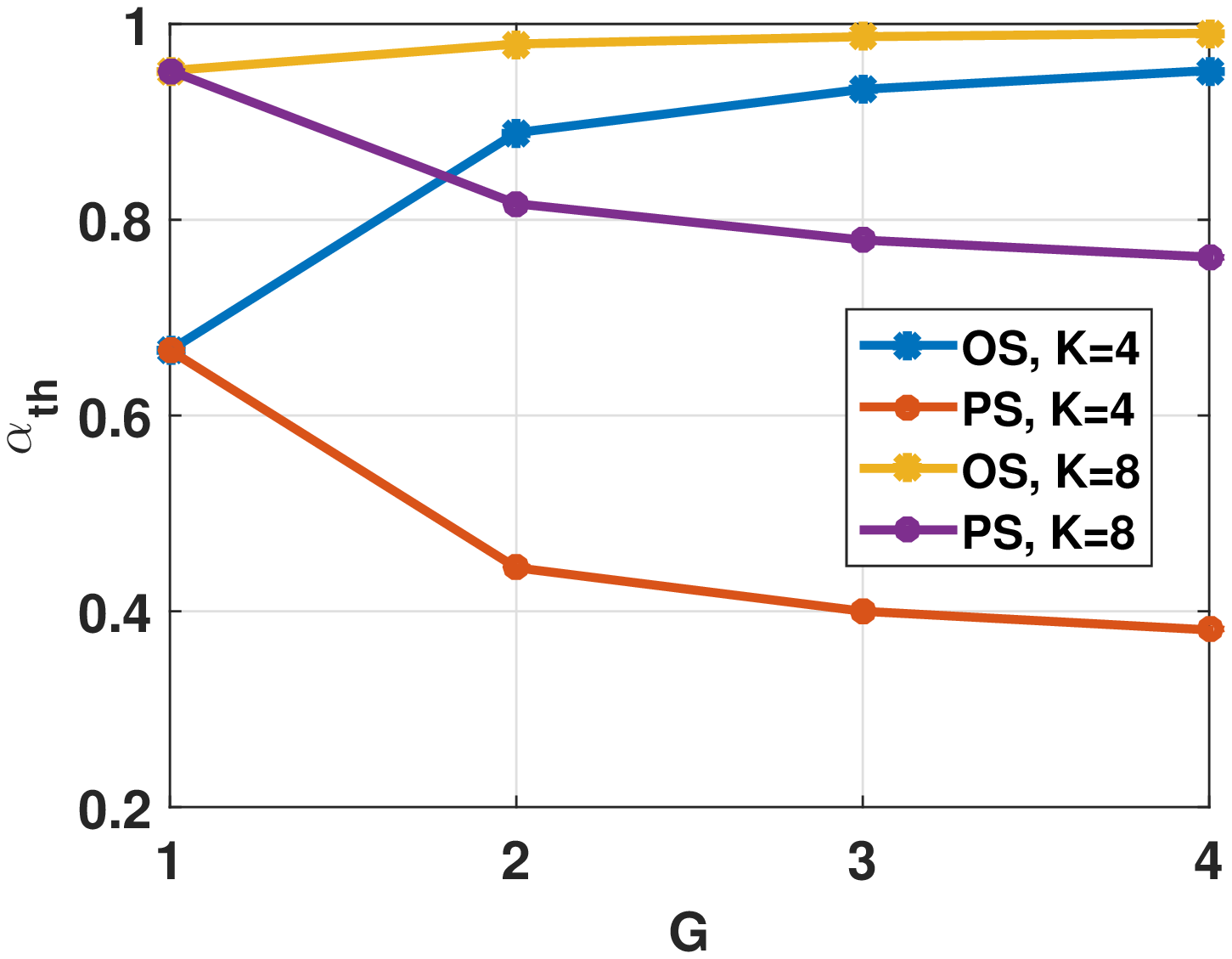}%
	}
	\subfloat[$\alpha_{\mathrm{th}}^{\mathrm{PS}}$ and $\alpha_{\mathrm{th}}^{\mathrm{OS}} $ with respect to $M$\label{sfig:figura_1b}]{%
		\includegraphics[width=.50\linewidth]{{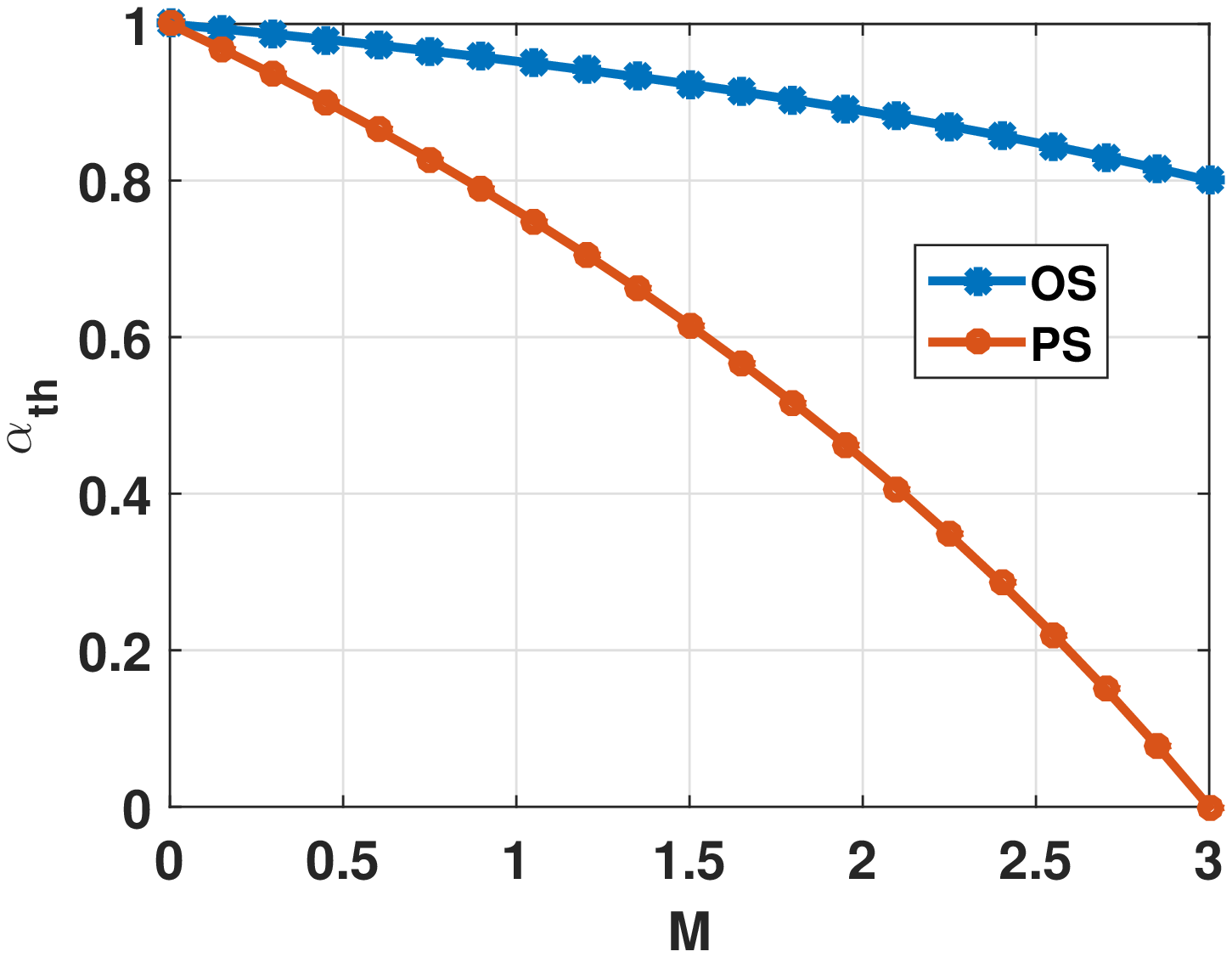}}%
	}
	\caption{CSIT qualities for the PS and OS to achieve $T_{\mathrm{D}}^{\mathrm{OS}}$}
	\label{}
\end{figure}

In this section, we numerically evaluate the PS.
First, we compare the \textit{delivery time} $T_{D}(\alpha)$ achieved under different CSIT qualities to the one achieved using the MAN scheme, i.e. $T_{\mathrm{D}}^{\mathrm{MAN}}$. This is given in Fig. \ref{fig:figura_0} for a setup with $K=8$ antennas, overloading factor $G=2$
and memories of size $Mf$ bits, with $M=1$.
We consider a library of $N_{\mathrm{f}}=K_{\mathrm{t}}$ throughout the simulations.
We observe that the PS leverages the available CSIT quality to reduce the \textit{delivery time} compared to MAN.
 In particular, for $\alpha=1$, the \textit{delivery time} is four times lower.
Next, we compare the PS with OS in terms of the minimum CSIT quality needed to achieve $T_{\mathrm{D}}^{\mathrm{OS}}$ in (\ref{OR_TD}).
In Fig. \ref{sfig:figura_1a}  we plot $\alpha_{\mathrm{th}}^{\mathrm{PS}}$ and $\alpha_{\mathrm{th}}^{\mathrm{OS}}$ as a function of $G$.
We consider a setup with $K=4,8$ and $M=2$. As we can see, the OS suffers when $G$ increases, while the PS, in addition to outperforming OS, benefits from a larger $G$.
Finally, in Fig. \ref{sfig:figura_1b} we plot $\alpha_{\mathrm{th}}^{\mathrm{PS}}$ and $\alpha_{\mathrm{th}}^{\mathrm{OS}}$ as a function of $M$ with $K=4$ and $G=2$. We can see that $\alpha_{\mathrm{th}}^{\mathrm{PS}}$ decreases much faster
than $\alpha_{\mathrm{th}}^{\mathrm{OS}}$ for increasing $M$.

\section{Conclusion} \label{conclusion}
In this paper, we studied the complementary gain of coded caching and spatial multiplexing gain for an overloaded
MISO BC. We investigated how to simultaneously combine the gain offered from the aggregate {global cache memory}, with the gain offered by the presence of available CSIT.
We introduced a novel scheme which superimposes Maddah-Ali Niesen scheme, to exploit the coded caching gain, to a multiuser transmission scheme, to leverage the available CSIT.  Analytical and numerical results showed a significant gain compared to the existing schemes, both in terms of \textit{delivery time} and CSIT quality requirement.

\section*{Appendix}

\subsection{Proof of Proposition 1}
In order to prove Proposition \ref{theo_1}, we need to evaluate the minimum of the function $T^*_{\mathrm{D},\eta}$ in (\ref{T_d,eta}) with respect to $\eta$ over the interval $ [\Gamma,K_{\mathrm{t}}] \cap \mathbb{Z} $, and we remember that
$K_{\mathrm{t}}=GK$.
The minimum value is indicated as $T^*_{\mathrm{D}}$ in (\ref{eq_T}).
We start by a relaxation, where $\eta$ is allowed to take any value in $[\Gamma,K_{\mathrm{t}}]$.
We can notice that $T^*_{\mathrm{D},\eta}$ is convex over such interval and $T^*_{\mathrm{D},K-1}=T^*_{\mathrm{D},GK}$,
hence $T^*_{\mathrm{D},\eta}$ has a minimum in \mbox{$[K-1,K_{\mathrm{t}}]$}.
The first derivative of $T^*_{\mathrm{D},\eta}$ with respect to $\eta$ is equal to zero for $\eta=x$, given in (\ref{x}).
By restricting $\eta$ to be an integer, i.e. $\eta \in [\Gamma,K_{\mathrm{t}}] \cap \mathbb{Z}$, we have that  $T^*_{{\mathrm{D}},\eta}$ is minimum in either $\floor{x}$ or $\ceil{x}$, from which the expression  in (\ref{UP}) follows.
It can be easily verified that $T^*_{\mathrm{D},K-1}=T^*_{\mathrm{D},GK}=T^{\mathrm{OS}}_{\mathrm{D}}$, where $T^{\mathrm{OS}}_{\mathrm{D}}$ is given in (\ref{OR_TD}). Hence, $T^*_{\mathrm{D}} < T^{\mathrm{OS}}_{\mathrm{D}} \leq T^{\mathrm{MAN}}_{\mathrm{D}}$.

\subsection{Proof of Proposition 2}

In order to prove Proposition \ref{theo_2}, we need to find the value of $T_{\mathrm{D}}(\alpha)$ in (\ref{eq_T_alpha}).
We have
\begin{equation}
T_{\mathrm{D}}(\alpha)= \min_{\eta' \in [\Gamma,K_{\mathrm{t}}] \cap \mathbb{Z}} T_{\mathrm{D}, \eta'}(\alpha)
\end{equation}
where $T_{\mathrm{D}, \eta'}(\alpha)$ is defined as in (\ref{formula_1}).
As explained in \mbox{Section \ref{csit_relaxation}}, in case $\alpha \geq \alpha^*$, it is possible to achieve the same \textit{delivery time}
as in the case of perfect CSIT. Hence, $T_{\mathrm{D}}(\alpha)=T_{\mathrm{D}}(1)=T_{\mathrm{D}}^*$.
We now focus on the case $\alpha < \alpha^*$.
We start by considering $\alpha_{\eta'}^{*}$ in Section \ref{imperfect_delivery_time}
\begin{equation}
\alpha_{\eta'}^{*}=\frac{Q_{\eta'}^{\mathrm{(p)}}/K}{Q_{\eta'}^{\mathrm{(p)}}/K+Q_{\eta'}^{\mathrm{(c)}}}=\frac{1}{1+K{Q^{(\mathrm{c})}_{\eta'}}/{Q^{(\mathrm{p})}_{\eta'}}}.
\end{equation}
It can be shown that $Q_{\eta'}^{\mathrm{(c)}}$ is a decreasing function of $\eta'$ while $Q_{\eta'}^{\mathrm{(p)}}$ is an increasing function of $\eta'$. It follows that $\alpha_{\eta'}^{*}$ is also an increasing function of $\eta'$.
Then, by denoting the total number of files to be transmitted as $Q_{\eta'}$, we have
\begin{equation}
Q_{\eta'}=Q^{(\mathrm{p})}_{\eta'}+Q^{(\mathrm{c})}_{\eta'}=K_{\mathrm{t}}-\frac{M}{N_{\mathrm{f}}} \frac{1}{\eta'+1}(K^2_{\mathrm{t}}+K_{\mathrm{t}})
\end{equation}
which is also an increasing function of  $\eta'$.
Moreover, from (\ref{formula_1}), the following relationship is verified
\begin{equation}
Q_{\eta'}=T^*_{\mathrm{D},\eta'}(1+(K-1)\alpha^*_{\eta'}).
\end{equation}
Given that we are considering $\alpha < \alpha^*$ and we know that $\alpha_{\eta^*}^*=\alpha^*$ and $\alpha_{\Gamma}^*=0$, there must exist an  $\eta$ such that \mbox{$\eta= \arg \max_{\eta' \in [\Gamma,\eta^*) \cap \mathbb{Z}} \{\eta':\alpha_{\eta'}^* \leq \alpha \}$}.
Moreover, from the proof of Proposition 1, we know that $T^*_{\mathrm{D},\eta'}$ is a decreasing function with respect to $\eta'$ in $[\Gamma,\eta^*) \cap \mathbb{Z}$.
From (\ref{formula_1}), considering \mbox{$\eta' \leq \eta$}, we have \mbox{$T_{\mathrm{D},\eta'}(\alpha) = T_{\mathrm{D},\eta'}^* \geq T_{\mathrm{D},\eta}^* $}.
In case of $\eta' \geq \eta+1$, we have
\begin{equation}
T_{\mathrm{D},\eta'}(\alpha)=\frac{Q_{\eta'}^{(\mathrm{p})}+Q_{\eta'}^{(\mathrm{c})}}{1+(K-1)\alpha} \geq \frac{Q_{\eta+1}^{(\mathrm{p})}+Q_{\eta+1}^{(\mathrm{c})}}{1+(K-1)\alpha} =T_{\mathrm{D},\eta+1}(\alpha).
\end{equation}
Hence,
$T_{\mathrm{D}}(\alpha)$ is the minimum between $T_{\mathrm{D},\eta}^*$ and $T_{\mathrm{D},\eta+1}(\alpha)$, from which Proposition 2 follows.
\subsection{Proof of Proposition 3}
From the proof of Proposition 1, we know that  $T^*_{\mathrm{D},\eta'}$ is a decreasing function with respect to $\eta'$ in $[\Gamma,\eta^*] \cap \mathbb{Z}$ and increasing in
$[\eta^*,K_{\mathrm{t}}] \cap \mathbb{Z}$.
Hence, there must exist
\begin{equation}
\eta= \arg \min_{\eta' \in [\Gamma ,\eta^*]  \cap \mathbb{Z}} \{\eta':T_{\mathrm{D},\eta'}^* \leq T \}
\end{equation}
and
\begin{equation}
\zeta= \arg \max_{\eta' \in [\eta^* ,K_{\mathrm{t}}]  \cap \mathbb{Z}} \{\eta':T_{\mathrm{D},\eta'}^* \leq T \}.
\end{equation}
In case of $\eta' < \eta$ or $\eta' > \zeta$, we have $T^*_{\mathrm{D},\eta'} > T$, thus  the \textit{delivery time} $T$ is not achievable.
On the other hand, by considering $\eta' \in [\eta,\zeta]$ and applying (\ref{formula_1}), the \textit{delivery time} $T$ is achieved by a CSIT quality given by
\begin{equation}
\alpha_{\eta'}= \frac{{Q_{\eta'}^{(\mathrm{p})}+Q_{\eta'}^{(\mathrm{c})}}-T}{(K-1)T} .
\end{equation}
As expressed in the proof of \mbox{Proposition 2}, \mbox{$Q_{\eta'}^{(\mathrm{p})}+Q_{\eta'}^{(\mathrm{c})}$} increases with $\eta'$.
It follows that \mbox{$\eta = \arg \min_{\eta' \in [\eta ,\zeta]  \cap \mathbb{Z}} \alpha_{\eta'}$}.
 From  \mbox{$Q_{\eta}^{(\mathrm{p})}+Q_{\eta}^{(\mathrm{c})}=T^*_{\mathrm{D},\eta}(1+(K-1)\alpha^*_{\eta})$}, we obtain (\ref{f_proposition_2}).

\ifCLASSOPTIONcaptionsoff
  \newpage
\fi

\bibliographystyle{ieeetr}
\bibliography{test}

\begin{thebibliography}{10}

\bibitem{Golrezaei2012}
N.~Golrezaei, K.~Shanmugam, A.~G. Dimakis, A.~F. Molisch, and G.~Caire,
  ``Femtocaching: Wireless video content delivery through distributed caching
  helpers,'' in {\em Proc. IEEE INFOCOM}, pp.~1107--1115, Mar. 2012.

\bibitem{maddahali_1}
M.~A. Maddah-Ali and U.~Niesen, ``Fundamental limits of caching,'' {\em IEEE
  Trans. Inf. Theory}, vol.~60, pp.~2856--2867, May 2014.

\bibitem{maddahali_2}
M.~A. Maddah-Ali and U.~Niesen, ``Decentralized coded caching attains
  order-optimal memory-rate tradeoff,'' {\em IEEE/ACM Trans. Net.}, vol.~23,
  pp.~1029--1040, Aug. 2015.

\bibitem{Gesbert2010}
D.~Gesbert, S.~Hanly, H.~Huang, S.~S. Shitz, O.~Simeone, and W.~Yu,
  ``Multi-cell {MIMO} cooperative networks: A new look at interference,'' {\em
  IEEE J. Sel. Areas Commun.}, vol.~28, pp.~1380--1408, Dec. 2010.

\bibitem{Jafar2011}
S.~A. Jafar, ``Interference alignment � a new look at signal dimensions in a
  communication network,'' {\em Found. Trends Commun. Inf. Theory}, vol.~7,
  no.~1, pp.~1--134, 2011.

\bibitem{Maddah-Ali2015}
M.~A. Maddah-Ali and U.~Niesen, ``Cache-aided interference channels,'' in {\em
  Proc. IEEE ISIT}, pp.~809--813, Jun. 2015.

\bibitem{Naderializadeh2016}
N.~Naderializadeh, M.~A. Maddah-Ali, and A.~S. Avestimehr, ``Fundamental limits
  of cache-aided interference management,'' in {\em Proc. IEEE ISIT},
  pp.~2044--2048, Jul. 2016.

\bibitem{elia_1}
J.~Zhang, F.~Engelmann, and P.~Elia, ``Coded caching for reducing
  {CSIT}-feedback in wireless communications,'' in {\em Proc. Allerton},
  pp.~1099--1105, Sep. 2015.

\bibitem{elia_2}
J.~Zhang and P.~Elia, ``Fundamental limits of cache-aided wireless {BC}:
  Interplay of coded-caching and {CSIT} feedback,'' {\em arXiv:1511.03961},
  2015.

\bibitem{kobayashi}
S.~Yang, K.~H. Ngo, and M.~Kobayashi, ``{Content delivery with coded caching
  and massive MIMO in 5G},'' in {\em Proc. ISTC}, pp.~370--374, Sep. 2016.

\bibitem{Piovano2016}
E.~Piovano, H.~Joudeh, and B.~Clerckx, ``Overloaded multiuser {MISO}
  transmission with imperfect {CSIT},'' {\em arXiv:1612.00628}, 2016.

\bibitem{jindal}
N.~Jindal, ``{MIMO} broadcast channels with finite-rate feedback,'' {\em IEEE
  Trans. Inf. Theory}, vol.~52, pp.~5045--5060, Nov. 2006.

\bibitem{Yang2013}
S.~Yang, M.~Kobayashi, D.~Gesbert, and X.~Yi, ``Degrees of freedom of time
  correlated {MISO} broadcast channel with delayed {CSIT},'' {\em IEEE Trans.
  Inf. Theory}, vol.~59, pp.~315--328, Aug. 2013.

\bibitem{Joudeh2016}
H.~Joudeh and B.~Clerckx, ``Sum-rate maximization for linearly precoded
  downlink multiuser {MISO} systems with partial {CSIT}: A rate-splitting
  approach,'' {\em IEEE Trans. Commun.}, vol.~64, pp.~4847--4861, Aug. 2016.

\end{thebibliography}


\begin{thebibliography}{1}







\bibitem{maddahali_1}
M. A. Maddah-Ali and U. Niesen, "Fundamental Limits of Caching," in \textit{IEEE Transactions on Information Theory}, vol. 60, no. 5, pp. 2856-2867, May 2014.

\bibitem{maddahali_2}
M. A. Maddah-Ali and U. Niesen, “Decentralized caching attains
order-optimal memory-rate tradeoff,” CoRR, vol. abs/1301.5848, 2013.
[Online]. Available: http://arxiv.org/abs/1301.5848

\bibitem{wang}
S. Wang, W. Li, X. Tian, and H. Liu, “Fundamental limits of
heterogenous cache,” CoRR, vol. abs/1504.01123, 2015. [Online].
Available: http://arxiv.org/abs/1504.01123

\bibitem{wang}
C.  Wang,  S.  H.  Lim,  and  M.  Gastpar,  “Information-theoretic caching: Sequential  coding  for  computing,”
CoRR, vol.  abs/1504.00553,  2015. [Online]. Available: http://arxiv.org/abs/1504.00553

\bibitem{yuang}
W.   Huang,   S.   Wang,   L.   Ding,   F.   Yang,   and   W.   Zhang,   “The
performance   analysis   of   coded   cache   in   wireless   fading   channel,”
CoRR
,  vol.  abs/1504.01452,  2015.  [Online].  Available:  http://arxiv.org/
abs/1504.01452

\bibitem{timo}
R.  Timo  and  M.  A.  Wigger,  “Joint  cache-channel  coding  over  erasure
broadcast   channels,”
CoRR
,   vol.   abs/1505.01016,   2015.   [Online].
Available: http://arxiv.org/abs/1505.01016

\bibitem{tuninetti}
K. Wan, D. Tuninetti, and P. Piantanida, “On caching with more users than files,” arXiv: 1601.063834v2 [cs.IT], Jan. 2016.

\bibitem{MA_IC}
M. A. Maddah-Ali and U. Niesen, “Cache-aided interference channels,”
in
Proceedings  of  the  IEEE  International  Symposium  on  Information
Theory (ISIT’2015)
, Hong-Kong, China, 201

\bibitem{elia_1}
J. Zhang, F. Engelmann and P. Elia, "Coded caching for reducing CSIT-feedback in wireless communications," \textit{2015 53rd Annual Allerton Conference on Communication, Control, and Computing (Allerton)}, Monticello, IL,

\bibitem{elia_2}
J. Zhang and P. Elia, “Fundamental limits of cache-aided wireless BC:
interplay of coded-caching and CSIT feedback,” arXiv, 1511.03961,
Apr., 2016 2015.

\bibitem{jindal}
N. Jindal, "MIMO Broadcast Channels With Finite-Rate Feedback," in \textit{IEEE Transactions on Information Theory}, vol. 52, no. 11, pp. 5045-5060, Nov. 2006. 2015, pp. 1099-1105.

\bibitem{bruno}
B. Clerckx, H. Joudeh, C. Hao, M. Dai and B. Rassouli, "Rate splitting for MIMO wireless networks: a promising PHY-layer strategy for LTE evolution," in \textit{IEEE Communications Magazine}, vol. 54, no. 5, pp. 98-105, May 2016.


\bibitem{Jafar_proof}
Arash Gholami Davoodi, Syed A. Jafar, "Aligned Image Sets under Channel Uncertainty: Settling Conjectures on the Collapse of Degrees of Freedom under Finite Precision CSIT," \textit{IEEE Transactions on Information Theory}.





\end{thebibliography}
\end{document}